\documentclass[12pt,a4paper,floatfix]{iopart}
\usepackage{amssymb,graphicx,cite}
\begin{document}
 

 


 
 
   
 
\title[Superfluid-Insulator Transition in a Condensate]{Dynamical
classical
superfluid-insulator transition in a
Bose-Einstein condensate on an optical lattice}
 
\author{Sadhan K. Adhikari}
\address{Instituto de F\'{\i}sica Te\'orica, Universidade Estadual
Paulista, 01.405-900 S\~ao Paulo, S\~ao Paulo, Brazil}

\date{\today}

\begin{abstract}

We predict a dynamical classical superfluid-insulator transition in a
Bose-Einstein condensate (BEC) trapped in a combined optical and
axially-symmetric  harmonic 
potentials initiated by a periodic modulation of the radial trapping
potential.  The transition is marked by a loss of 
phase coherence in the BEC  and a subsequent
destruction of the interference pattern upon free expansion. 
For a weak modulation of the radial  potential the phase coherence is
maintained. For a stronger modulation and a longer time of holding  in
the 
modulated trap, the  phase coherence is      destroyed signaling a
classical superfluid-insulator transition. The results are illustrated  
by a complete numerical solution of the   axially-symmetric mean-field
Gross-Pitaevskii equation for a  repulsive BEC. Suggestion for future
experiment is made.

\end{abstract}

\maketitle

The experimental loading of a cigar-shaped Bose-Einstein condensate 
(BEC) in  both one- \cite{1,2} and three-dimensional \cite{greiner}
optical
lattice potentials
has allowed the study of quantum phase
effects on a macroscopic scale such as interference of matter
waves \cite{kett1}.  There have been several theoretical studies on
different
aspects of a BEC in  one-  \cite{th} and  three-dimensional
\cite{adhi}
optical lattice
potentials.
The phase coherence between
different sites of a trapped BEC on an optical lattice has been
established in  recent experiments
\cite{1,2,cata,greiner,ari} through the formation of
distinct interference pattern when the traps are removed. 
In a
one-dimensional optical lattice potential the expanding pattern consists
of a large
central
piece and two smaller ones moving  in opposite
directions on a straight line \cite{cata}. In a two-dimensional  optical
lattice potential
the
pattern
consists of a large
central
piece and eight others on the sides of an expanding square \cite{adhi}. In
a
three-dimensional  optical lattice potential the 
pattern
consists of a large
central piece and twenty six others on the surface of an expanding cube
\cite{greiner,adhi}.

The interference pattern is a consequence of phase coherence in the BEC
generated due to  free quantum tunneling of atoms from one optical lattice
site to
another originating in the superfluid state of the system
\cite{greiner,stoof}. Equal phase at all points or a slowly (and orderly)
varying phase are 
 the ideal examples of
 coherent phase. On the other hand, a rapidly (or arbitrarily)  varying
phase in space is usually incoherent.

It has been demonstrated for  a
three-dimensional
optical trap potential 
by Greiner {\it et al.}
\cite{greiner} that, as the strength of the optical potential traps 
is increased, the quantum tunneling of condensed atoms from one optical
site to
another is
stopped resulting in a loss of superfluidity and phase coherence in the
BEC. Consequently, no interference pattern is formed upon free expansion
of such a BEC which is termed a Mott insulator state. This phenomenon
represents a superfluid to Mott insulator quantum phase transition. 
The phase on an optical lattice site and the number of atoms in that
site play the roles of conjugate variables obeying the Heisenberg
uncertainty principle of quantum mechanics \cite{stoof}.  In the
superfluid state the coherent phase is considered to be known and
consequently the number of atoms on each site is unknown thus allowing
a free movement of atoms from one site to another \cite{greiner}. In
the Mott insulator state the phase is entirely arbitrary across the
optical lattice sites  and the number of
atoms at each site is fixed and their free passage from one site to
another is stopped.
As the
strength of  the optical potential traps  in the  Mott insulator state is
reduced the superfluidity is restored in a short time via a  Mott
insulator to superfluid quantum phase transition. This reversible quantum
phase transition may occur at absolute zero (0 K) and is driven by
Heisenberg's uncertainty principle \cite{greiner} and not by thermal
fluctuations
involving  energy  
as in a 
classical phase  transition. As the temperature approaches absolute zero
all thermal fluctuations die out and at 0 K classical phase transitions
are necessarily 
excluded.

Following a suggestion by Smerzi {\it et al.} \cite{sm}, Cataliotti {\it
et al.} \cite{cata2} have demonstrated in a novel experiment the loss of
phase coherence and superfluidity in a BEC trapped in a one-dimensional
optical-lattice and harmonic potentials when the center of the harmonic
potential is suddenly displaced along the optical lattice through a
distance larger than a critical value.  Then a modulational instability
takes place in the BEC and it cannot reorganize itself quickly enough and
the phase coherence and superfluidity of the BEC are lost.  The loss of
superfluidity is manifested in the destruction of the interference
pattern upon free expansion. However, for displacements smaller than the
critical distance the BEC can reorganize itself and the superfluidity is
maintained \cite{cata,cata2}.  Distinct from the quantum phase transition
observed by Greiner {\it et al.} \cite{greiner}, this modulational
instability responsible for the superfluid-insulator transition is
classical in nature \cite{sm,cata2}.  This process is also different from
the Landau dissipation mechanism \cite{sm,cata4}, occurring when the fluid
velocity is greater than local speed of sound. When Landau instability occurs,
the system lowers energy by emitting phonons \cite{cata4}. 
The present classical dynamical 
transition can be well described \cite{adhi1,sm,cata4} by the mean-field
Gross-Pitaevskii (GP) equation \cite{8}.

The above modulational   instability   is not the  unique dynamical classical
process leading to a  superfluid-insulator transition. Many
other
classical
processes leading to a rapid movement in the condensate can lead to such a
transition \cite{adx}. The movement should be rapid enough so that the BEC
cannot
reorganize itself to evolve through   phase coherent states. In
 \cite{cata2} a rapid translation of the BEC through the optical
lattice sites leads to the destruction of phase coherence. 
Here we suggest
that a rapid oscillation of the BEC  may
also lead to  a
superfluid-insulator transition. The oscillation is initiated by a
periodic modulation of the magnetic trapping potential  $\sim \omega^2$ in
the radial direction 
via $\omega^2 \to \omega^2(1+A\sin(\Omega \tau ))$ where $\tau $ is time,
$A$ an
amplitude, $\omega$ is the radial trapping frequency, and $\Omega$
is the frequency of modulation.  Such modulation
of the trapping potential is known to generate resonant
(collective) excitations in
the BEC which have been studied both theoretically \cite{6}
and experimentally
\cite{3}
in the absence of an optical lattice
potential. The study of such excitations in the presence of an
optical
lattice
potential has just began \cite{3a}. Similar collective  excitation
generated by a periodic modulation of the atomic scattering 
length \cite{abd}
has also been shown to lead to a classical superfluid-insulator 
transition \cite{ska}.

In the quantum phase transition \cite{greiner} the Mott insulator state
has a perfectly smooth probability distribution (modulus of the wave
function) across the optical lattice sites whereas the phase of the wave
function across the optical lattice sites remains entirely arbitrary. In
the dynamical classical transition considered in this work, because of
classical oscillation of the BEC, the insulator state in the joint
traps
is marked by a
partially
disturbed (nonsmooth) probability distribution across the optical lattice
sites in
addition to the loss of phase coherence. However, the information about
the  destruction of superfluidity 
in both quantum and
classical cases is { \it not} solely contained in the initial  probability
distribution. Consequently,
the BEC needs to be released from the joint traps and the formation of the
interference pattern studied for a definite conclusion about the
destruction of superfluidity.

\newpage

As the present transition is classical or mean-field-type in nature,
we base the present study on the numerical solution of the
time-dependent mean-field
axially-symmetric GP equation \cite{8} in the presence
of a
combined 
harmonic and optical potential traps. 
The time-dependent BEC wave
function $\Psi({\bf r};t)$ at position ${\bf r}$ and time $t $
is described by the following  mean-field nonlinear GP equation
\cite{8}
\begin{eqnarray}\label{a} \left[- i\hbar\frac{\partial
}{\partial t}
-\frac{\hbar^2\nabla^2   }{2m}
+ V({\bf r})
+ gN|\Psi({\bf
r};t)|^2
 \right]\Psi({\bf r};t)=0,
\end{eqnarray}
where $m$
is
the mass and  $N$ the number of atoms in the
condensate,
 $g=4\pi \hbar^2 a/m $ the strength of interatomic interaction, with
$a$ the atomic scattering length.  In the presence of the combined
axially-symmetric and optical lattice traps 
     $  V({\bf
r}) =\frac{1}{2}m \omega ^2(\rho ^2+\nu^2 z^2) +V_{\mbox{opt}}$ where
 $\omega$ is the angular frequency of the harmonic trap 
in the radial direction $\rho$,
$\nu \omega$ that in  the
axial direction $z$, with $\nu$ the aspect ratio, and $V_{\mbox{opt}}$ is
the optical lattice trap introduced later.  
The normalization condition  is
$ \int d{\bf r} |\Psi({\bf r};t)|^2 = 1. $

In the axially-symmetric configuration, the wave function
can be written as 
$\Psi({\bf r}, t)= \psi(\rho ,z,t)$.
Now  transforming to
dimensionless variables $\hat \rho =\sqrt 2 \rho /l$,  $\hat z=\sqrt 2
z/l$,
$\tau=t
\omega, $
$l\equiv \sqrt {\hbar/(m\omega)}$,
and
${ \varphi(\hat \rho,\hat z;\tau)} \equiv   \hat \rho \sqrt{{l^3}/{\sqrt
8}}\psi(\rho ,z;t),$   (\ref{a}) becomes \cite{11}
\begin{eqnarray}\label{d1}
&\biggr[&-i\frac{\partial
}{\partial \tau} -\frac{\partial^2}{\partial
\hat \rho^2}+\frac{1}{\hat \rho}\frac{\partial}{\partial \hat \rho}
-\frac{\partial^2}{\partial
\hat z^2}
+\frac{1}{4}\left(\hat \rho^2+\nu^2 \hat z^2\right) \nonumber \\
&+&\frac{V_{\mbox{opt}}}{\hbar \omega} -{1\over \hat \rho^2}  +                                                          
8\sqrt 2 \pi n\left|\frac {\varphi({\hat \rho,\hat z};\tau)}{\hat
\rho}\right|^2
 \biggr]\varphi({ \hat \rho,\hat z};\tau)=0, 
\end{eqnarray}
where nonlinearity
$ n =   N a /l$. In terms of the 
one-dimensional probability 
 $P(z,t) \equiv 2\pi$ $\int_0 ^\infty 
d\hat \rho |\varphi(\hat \rho,\hat z,\tau)|^2/\hat \rho $, the
normalization
of
the
wave
function 
is given by $\int_{-\infty}^\infty d\hat z P(z,t) = 1.$  
The probability 
$P(z,t)$ is  useful in the study of the present problem under the
action of the optical lattice potential, specially in the
investigation of the formation and
evolution of the interference pattern after the removal of the
trapping potentials.

In the  experiment of Cataliotti {\it et al.} \cite{cata}
with repulsive $^{87}$Rb atoms in the hyperfine state $F=1,
m_F=-1$, 
the radial trap frequency was  $ \omega =
2\pi \times 92$ Hz. The
optical
potential created with the standing-wave laser field of wavelength 
$\lambda=795$ nm is given by $V_{\mbox{opt}}=V_0E_R\cos^2 (k_Lz)$,
with $E_R=\hbar^2k_L^2/(2m)$, $k_L=2\pi/\lambda$, and $V_0$ $ (<12)$ the 
strength. For the mass $m=1.441\times 10^{-25}$ kg  of $^{87}$Rb the
harmonic
oscillator length $l=\sqrt {\hbar/(m\omega)} = 1.126$ $\mu$m and 
and the 
present
dimensionless time unit  $\omega ^{-1} =
1/(2\pi\times 92)$ s $=1.73$ ms. In terms of the dimensionless laser wave
length $\lambda _0= \sqrt2\lambda/l \simeq 1$, the dimensionless 
standing-wave energy parameter $E_R/(\hbar \omega)= 4\pi^2/\lambda _0^2$.
Hence in 
dimensionless unit $V_{\mbox{opt}}$ of 
  (\ref{d1}) is
\begin{equation}\label{pot}
\frac{ V_{\mbox{opt}}}{\hbar \omega}=V_0\frac{4\pi^2}{\lambda_0^2} 
\left[
\cos^2 \left(
\frac{2\pi}{\lambda_0}\hat z
\right)
 \right].
\end{equation}
Although we employ the dimensionless space units $\hat \rho$ and $\hat z$
and
time unit
$\tau$ in numerical calculation, the results are reported in actual units 
$r$ $\mu$m, $z$ $\mu$m  and $t$ ms. In the conversion we used the
parameters of the experiment of Cataliotti {\it et al.} \cite{cata}, e.g.,
$\rho= 0.8\hat \rho$ $\mu$m, $z=0.8\hat z$ $\mu$m, and $t=1.73\tau$ ms.

We solve   (\ref{d1}) numerically  using a   
split-step time-iteration
method
with  the Crank-Nicholson discretization scheme described recently
\cite{11x}.  
The time iteration is started with the  harmonic oscillator solution
of   (\ref{d1}) with
 $n=0$: $\varphi(\hat \rho,\hat z) = [\nu
/(8\pi^3)  ]^{1/4}$
$\hat \rho e^{-(\hat \rho^2+\nu \hat z ^2)/4}$ 
\cite{11}. 
The
nonlinearity $n$  and  the optical lattice potential parameter $V_0$ 
are  slowly increased by equal amounts in $10000n$ steps of 
time iteration until the desired value of $n$ and  $V_0$        are
attained. Then, without changing any
parameter, the solution so obtained is iterated 50 000 times so that a
stable
solution  is obtained 
independent of the initial input
and time and space steps.

The one-dimensional pattern of  BEC  on the
optical
lattice  for a specific nonlinearity and the 
interference pattern upon free expansion of such a BEC have been
recently studied using the numerical solution of
 (\ref{d1}) \cite{adhi1}. Here we study the destruction of this
interference pattern after the application of a periodic modulation of the
radial trapping potential in  (\ref{d1}) via 
\begin{equation}\label{rep}
\hat \rho^2/4  \to (\hat \rho^2/4 ) [1+A\sin (\Omega \tau)],
\end{equation}
while the axial trapping potential is left unchanged \cite{6,3}.  In the
present
model
study we employ  nonlinearity $n=5$,  the axial trap parameter 
$\nu =0.5$, 
and   the optical lattice strength 
$V_0=6$ throughout. First we calculate  the ground-state wave
function in the combined harmonic  and optical lattice potentials.

Modulation (\ref{rep}) of the trapping potential may lead to resonant
oscillation of the BEC and such resonances have been studied in the case
of harmonic trap alone \cite{6}. For a very small $A$, prominent
resonances
appear in the BEC oscillation when the modulation frequency $\Omega$ is an
integral
multiple ${\cal N}$
of the harmonic oscillator frequency  $\omega$  ($\Omega={\cal N}\omega$)  
\cite{6}. This is quite expected from our wisdom in linear classical
physics where resonances appear when the driving frequency is a multiple
of the characteristic frequency of oscillation.  Actually, for a finite
$A$, resonances appear for a band of modulation frequency \cite{6}
($\Omega ={\cal
N}\omega\pm \Delta_{\cal N}$) where $\Delta_{\cal N}$ defines the spread
of frequency values. The resonance becomes more prominent with the
increase of the parameter $A$ or ${\cal N}$.  At resonance the BEC
executes rapid oscillation which is responsible for 
the destruction of superfluidity in the BEC via a classical
dynamical transition provided that the 
time of stay of the BEC in
the modulated
magnetic trap, called hold time,  is larger than
a critical value.  A large value of $A$ and/or $\Omega$ facilitates the
dynamical classical superfluid-insulator transition. 
We illustrate this
fact in the
following for ${\cal
N}=1$ and 2. A careful study of the present phenomenon will aid in the
understanding of resonance in nonlinear physics. Although the generation
of resonance in different linear problems is well understood, the same in
nonlinear physics is just starting.

\begin{figure}[!ht]
 
\begin{center}

\includegraphics[width=.8\linewidth]{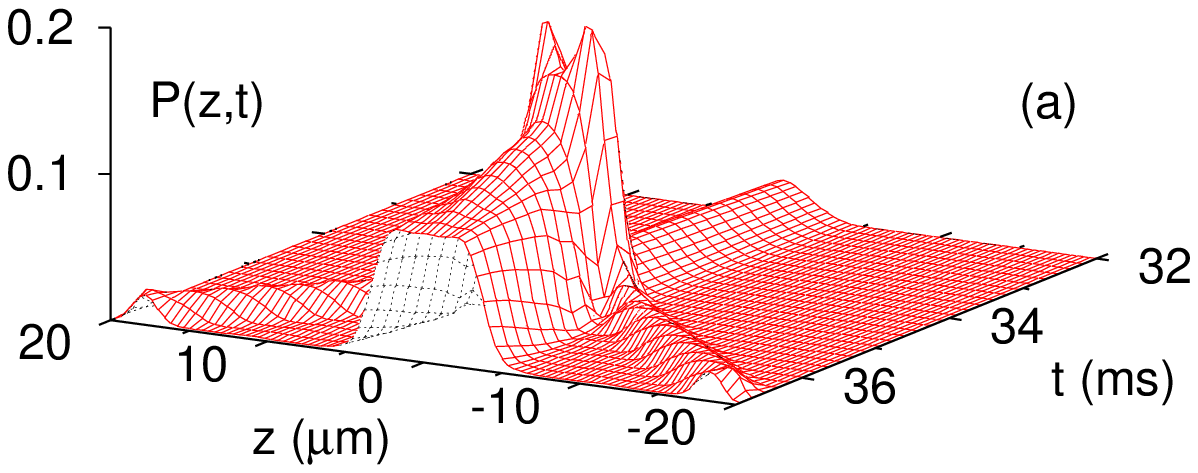}
\includegraphics[width=.8\linewidth]{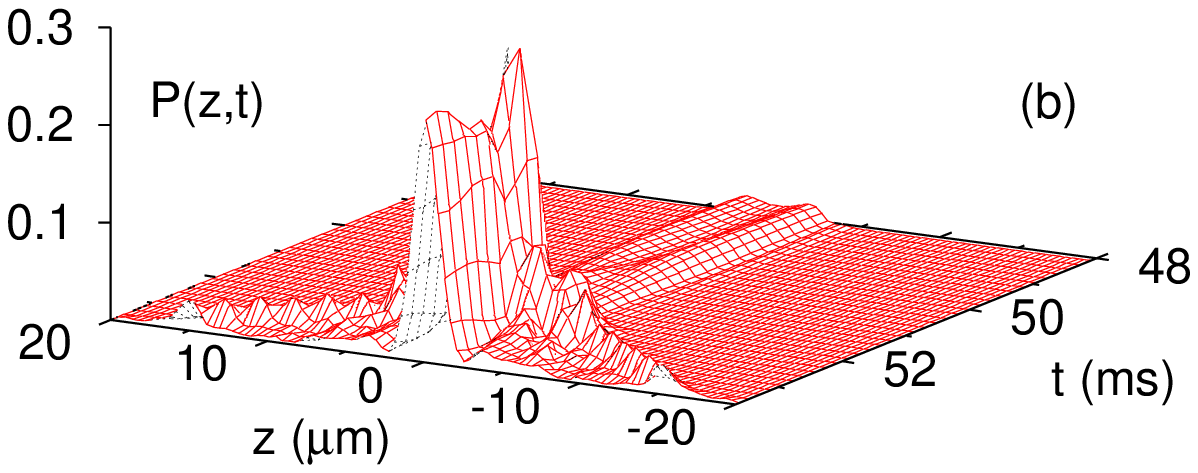}
\includegraphics[width=.8\linewidth]{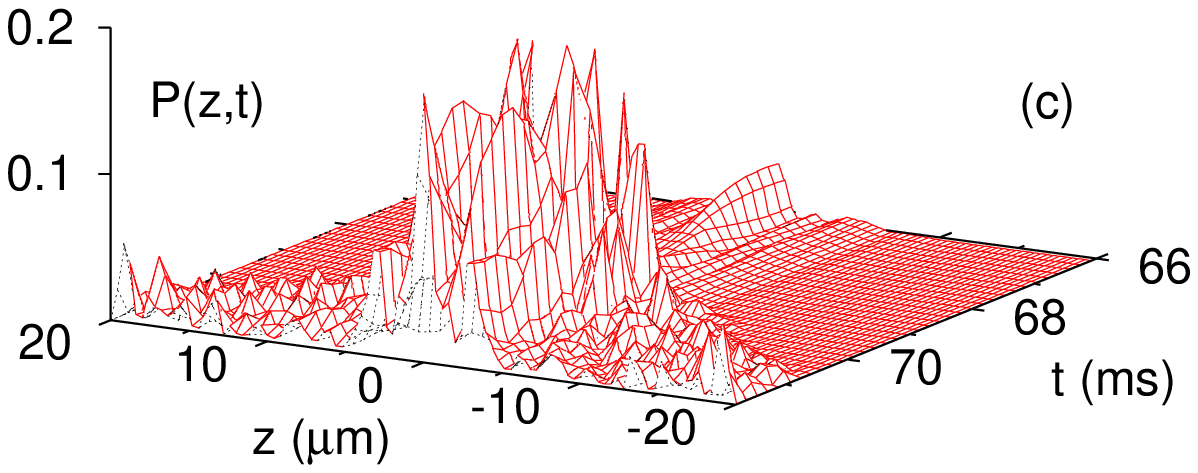}
\end{center}
 
\caption{One-dimensional probability  $P(z,t)$
vs. $z$ and $t$ for the BEC on optical lattice 
under the action of modulation (\ref{rep}) with  $\Omega
=\omega$
and $A=0.5$ 
in the radial magnetic trap 
and upon the removal of the
combined traps after hold times (a) 35 ms, (b) 52 ms, and (c) 69 ms,  in
the modulated   radial trap.  
} \end{figure}

As the present calculation is performed with the full wave function
without approximation, phase coherence among different wells of the 
optical lattice   is automatically guaranteed in the initial state. As a
result when
the
condensate is released from the combined trap, a matter-wave  
interference
pattern is
formed in a few milliseconds as described in  \cite{adhi1}.  The atom
cloud released from one lattice
site expand, and overlap and interfere with atom clouds from neighboring
sites to form the robust interference pattern. 
The pattern consists of a central peak 
and two symmetrically spaced peaks, each containing about $10\%$ of 
total number of atoms
moving in opposite directions \cite{cata}.

Next we consider an oscillating BEC in the combined harmonic 
and  optical traps. If we introduce the modulation of the radial trapping
potential (\ref{rep}) 
after the
formation of the BEC in the
combined trap, the condensate will be out of equilibrium and start to
oscillate. As the height of the potential-well barriers on the optical
lattice  
is much larger than the energy of the system, the atoms in the condensate
will move  by tunneling through the potential barriers.
This fluctuating transfer of Rb atoms across the potential barriers
is due to Josephson effect in  a neutral quantum liquid \cite{cata}.
We demonstrate 
that 
the phase coherence between
different wells of the condensate can be destroyed  during this process 
for rapid  oscillations with large amplitude and/or frequency
and  no  matter-wave interference pattern will be formed after the
removal
of
the joint traps.

\begin{figure}[!ht]
 
\begin{center}

\includegraphics[width=.8\linewidth]{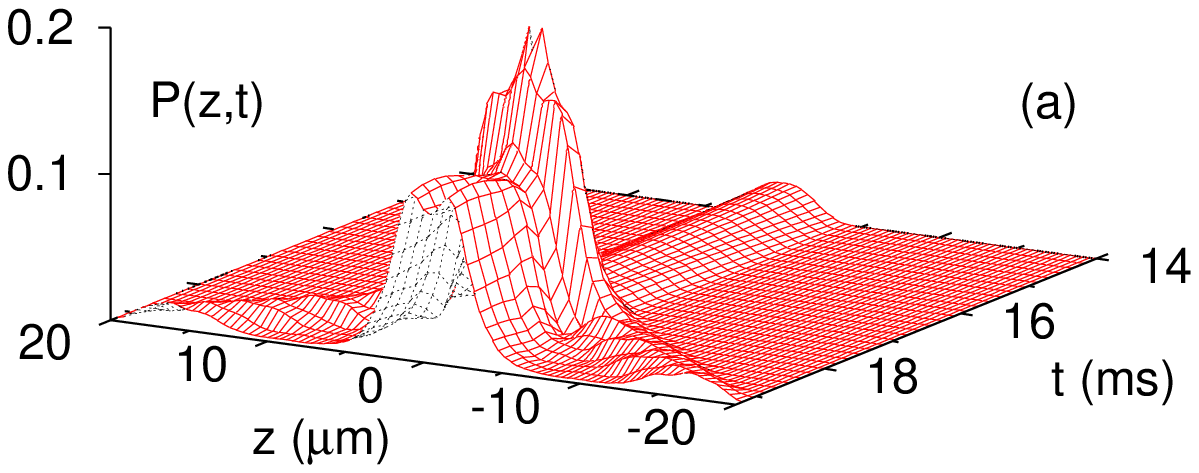}
\includegraphics[width=.8\linewidth]{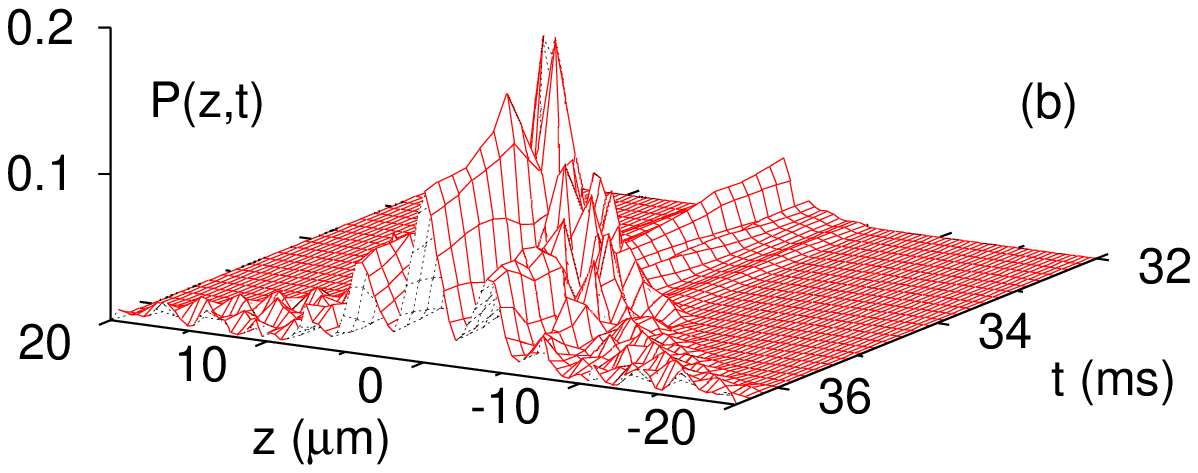}
\end{center}
 
\caption{One-dimensional probability  $P(z,t)$
vs. $z$ and $t$ for the BEC on optical lattice 
under the action of modulation (\ref{rep}) with  $\Omega
=2\omega $
and $A=0.5$ 
in the radial magnetic trap 
and upon the removal of the
combined traps after hold times (a) 17 ms, and (b) 
35 ms in the modulated   radial trap.  
} \end{figure}

Now  we explicitly study  the destruction of superfluidity in the
condensate upon the application of  modulation (\ref{rep})
when 
the BEC is allowed to stay in this modulated  trap for a certain  interval
of time more than a critical value
(hold time). For small  $A$ and $\Omega$ away from resonance, the BEC
executes slow
oscillation maintaining the phase coherence.  For large  $A$ and $\Omega$ 
and near resonance,
the BEC executes rapid oscillation 
\cite{3,3a} which 
 results in a destruction of superfluidity.
The destruction of  superfluidity  for a larger hold
time in the modulated  trap manifests 
in the disappearance of the interference pattern upon free expansion which
can be studied
experimentally.

For numerical simulation we allow the BEC to evolve on  a lattice with 
$r \le 15$ $\mu$m and 20 $\mu$m $\ge z \ge$ $-20$ $\mu$m after the
modulation (\ref{rep}) with $\Omega =\omega$ and $A=0.5$ is applied 
and study the
the system after different   hold times.
The 
probability densities $P(z,t)$
 are plotted in  figures  1 (a), (b)   and (c), 
for hold times  35 ms, 52 ms, and 69 ms,  respectively. For the hold
time of 35 ms prominent interference
pattern is
formed upon free expansion as we can see in figure  1 (a). The
interference 
pattern is slowly destroyed as hold time is increased as  we can see in
 figures  1 (b)  and (c). In  figure  1(a)  
three separate pieces in the  interference pattern  
corresponding  to three distinct trails can be identified. 
However,
as the hold time in the displaced trap increases the maxima
of the interference pattern
mixes up and finally for the  hold time of 
69 ms  the interference pattern is
completely destroyed as we find in  figure  1 (c).

As the BEC is allowed to evolve for a substantial  interval of time after
the  modulation (\ref{rep})  of the radial trapping potential is applied,  
a
dynamical instability of classical nature sets in which destroys
superfluidity \cite{sm,cata2}. This has been
explicitly demonstrated in the present simulation which results in the
destruction of the interference pattern.

The destruction of superfluidity  is facilitated as the amplitude $A$ or
frequency $\Omega $ of the radial modulation (\ref{rep}) is increased. 
We demonstrate this for an increase in  $\Omega $ in the following.
In  figures  2 (a) and (b) we present the evolution of probability 
$P(z,t)$ after the application of the  modulation  with $A= 0.5$ and 
$\Omega =2\omega$. With the increase of $\Omega $ from $\omega$ to
$2\omega$, the
destruction
of   superfluidity is facilitated as one can find from  figures  2. The
 superfluidity is destroyed for a hold time of 35 ms for $\Omega =2\omega$
(figure  2),
whereas it is maintained for the same hold time for  $\Omega =\omega$
(figure  1). An increase in the value of the parameter $A$ also increases
the resonant oscillation and we verified that the
destruction
of  superfluidity is also facilitated in the process. However, we do
not   present that study here.

In conclusion, using the  explicit numerical
solution
of the   GP equation 
we have studied in detail the destruction of  superfluidity
in a cigar-shaped condensate loaded in a combined axially-symmetric
harmonic  and  optical lattice traps  upon the application of 
a modulation of the radial trapping potential near resonance.   In the
absence of 
modulation,  the
formation of the interference pattern 
upon the removal of the combined
traps clearly demonstrates the
phase coherence \cite{adhi,adhi1}. 
The  superfluidity is maintained for a  slow  modulation
(\ref{rep})
of the radial trapping potential away from resonance. For  rapid
modulation of large 
amplitude and/or frequency near resonance, there  is
 a superfluid-insulator classical dynamical   transition,
provided
that 
the BEC is kept in the modulated trap for a certain hold time.
Consequently,
after release from the combined trap no interference pattern is formed.
For smaller
amplitude $A$ and/or frequency $\Omega$ of modulation (\ref{rep}) near
resonance, 
the superfluid-insulator
transition occurs for a larger hold time and vice versa. 
It is possible to study this novel phenomenon experimentally and compare
with the present theoretical prediction.

\ack  

The work was supported in part by the CNPq and FAPESP
of Brazil.

 \end{document}